\documentclass[prd,superscriptaddress,amsmath,notitlepage,onecolumn,12pt]{revtex4-1}
\usepackage{graphicx}
\usepackage{float}
\usepackage{epstopdf,cancel}
\usepackage{epsf,latexsym,bbm,euscript}
\usepackage{amssymb,amsmath}
\usepackage{mathtools} 
\usepackage{times,graphics}
\usepackage{soul,xcolor}
\usepackage{mathtools}
\usepackage[normalem]{ulem}
\newcommand{\D}{\mathrm{d}}
\newcommand{\p}{\partial}

\def\6{{\langle}}
\def\9{{\rangle}}

\newcommand{\be}{\begin{equation}}
\newcommand{\ee}{\end{equation}}
\newcommand{\ba}{\begin{eqnarray}}
\newcommand{\ea}{\end{eqnarray}}

\usepackage{xtab,afterpage}
\usepackage{hyperref}
\usepackage{setspace,calc}

\begin{document}

\title{Schr\"odinger's Black Hole Cat}
 
\author{Joshua Foo\footnote{Corresponding Author}}
\email{joshua.foo@uqconnect.edu.au}
\affiliation{Centre for Quantum Computation \& Communication Technology, School of Mathematics \& Physics, The University of Queensland, St.~Lucia, Queensland, 4072, Australia}

\author{Robert B.\ Mann}
\email{rbmann@uwaterloo.ca}
\affiliation{Department of Physics and Astronomy, University of Waterloo, Waterloo, Ontario N2L 3G1, Canada}
\affiliation{Perimeter Institute for Theoretical Physics, Waterloo, Ontario N2L 6B9, Canada}

\author{Magdalena Zych}
\email{m.zych@uq.edu.au}
\affiliation{Centre for Engineered Quantum Systems, School of Mathematics and Physics, The University of Queensland, St. Lucia, Queensland, 4072, Australia}

\begin{abstract}
\vspace*{1mm}
In the absence of a fully-fledged theory of quantum gravity, we propose a ``bottom-up'' framework for exploring quantum-gravitational physics by pairing two of the most fundamental concepts of quantum theory and general relativity, namely quantum superposition and spacetime. We show how to describe such ``spacetime superpositions'' and explore effects they induce upon quantum matter. Our approach capitalises on standard tools of quantum field theory in curved space, and allows us to calculate physical observables like transition probabilities for a particle detector residing in curvature-superposed de Sitter spacetime, or outside a mass-superposed black hole. Crucially, such scenarios represent genuine quantum superpositions of spacetimes, in contrast with superpositions of metrics which only differ by a coordinate transformation and thus are not different according to general relativity. 
\end{abstract}

\maketitle

\vspace{30mm}

\textit{Essay written for the Gravity Research Foundation 2022 Awards for Essays on Gravitation.}

\vspace{15mm}

\text{\hfill \hspace*{104mm} Submitted on March 31, 2022}

\vspace{5mm}

\newpage

A unified theory of quantum gravity continues to elude us despite numerous attempts over the last century to construct one.  The dynamical character of the spacetime metric obstructs the implementation of standard approaches to its quantization. Top-down attempts at unification such as string theory \cite{gubser1998gauge,seiberg1999string,polchinski1998string,witten1995string} and loop quantum gravity \cite{rovelli2008loop,rovelli2014covariant,thiemann2003lectures} have yet to succeed, raising questions as to the necessity or well-posedness of quantum gravity \cite{dyson2014graviton,garay1995quantum,carlip2008quantum}.

We propose taking a ``bottom-up" perspective on this problem, one that allows exploration of quantum gravity effects that do not require the full dynamical theory of quantum gravity.  Specifically, we focus on the most signature feature of quantum physics: superposition.

If indeed relativistic gravitation has a quantum description, it must be meaningful to consider a spacetime metric in a genuine quantum superposition. Any theory combining  spacetime and quantum theory should admit as a special case superpositions of semiclassical geometries. Recent experimental proposals attempting to gravitationally induce entanglement between two mesoscopic particles each in a spatial superposition~\cite{bose2017spin,marletto2017gravitationally}, if successful, have been argued to provide evidence
for superpositions of semiclassical spacetime geometries~\cite{christodoulou2019possibility}. Other studies have argued for inevitable decoherence mechanisms in systems involving macroscopically superposed gravitational sources (e.g.\ black holes and dark matter \cite{Arrasmith_2019,Demers_1996,allaliPhysRevLett.127.031301}). Rapid advances in tabletop techniques \cite{bosePhysRevA.105.032411,howlPRXQuantum.2.010325,carneyPRXQuantum.2.030330} show that such questions are not only of fundamental, but also practical interest, especially when  considering analog experiments, for example with Bose-Einstein condensates~\cite{Steihnauer2019Nature, barcelo2021superposing}.

The common feature of these investigations is that they focus on effects produced by spatial superpositions of massive objects. Such scenarios involve classical metrics that differ only by a coordinate transformation and are thus diffeomorphic. One can equivalently describe these ``spacetime superpositions'' in terms of a single classical spacetime where quantum systems residing within are prepared and measured in appropriate quantum states~\cite{zych2018relativity, Foo_2021} (we present this  argument further in this essay for completeness). 

Our new approach {admits  superposition of} genuinely different  spacetimes, i.e.\ superpositions of unique solutions of Einstein's field equations, which are thus not diffeomorphic. It can of course also describe superpostions of spacetimes  differing by a mere coordinate transformation, such as those arising from the above mentioned superpositions of locations of a source mass. {It allows us to}  explicitly show how and why the latter elicit the same effects as the corresponding quantum states of matter on a fixed background in a concrete scenario.

We do not propose a full quantum-gravitational theory for the emergence of quantum  spacetimes, but rather develop an operational framework describing dynamics of quantum systems residing within them. The framework is based on the standard paradigm of quantum fields in curved spacetime and quantum matter coupled to these fields. We are predominantly interested in the effects produced by spacetime superpositions on the quantum matter system(s), henceforth called detectors or probes, and the new insights they may engender. These quantum probes are modelled as 2-level systems linearly coupled to quantum fields, otherwise known as Unruh-deWitt (UdW) detectors~\cite{unruh1984happens}. This model has been utilized widely in relativistic and gravitational settings to investigate properties of quantum fields, vacuum entanglement,  black holes, and causal structure \cite{ng2014unruh,ng2017over,ng2018new,henderson2019btz,hodgkinson2012static,hodgkinson2012often,henderson2020quantum}. 

Before proceeding towards more complex situations involving superpositions of gravitational sources, it is instructive to first consider flat space analogues to such scenarios. For example, consider placing a UdW detector on a classical uniformly accelerated (Rindler) trajectory. Its response to the field (i.e.\ its transition probability) is a Planckian distribution with an associated temperature proportional to the magnitude of its acceleration, the well-known Unruh effect. This effect has a direct correspondence with the thermalization of the Hartle-Hawking vacuum outside a Schwarzschild black hole, which will likewise cause a detector to ``thermalize'' to the Hawking temperature, inversely proportional to the black hole's mass. 

Let us now place the detector in a 
superposition of semiclassical trajectories with different uniform accelerations \cite{foo2020unruhdewitt,barbado2020unruh,foo2020thermality}:
\begin{align}
    | \psi \rangle &= \frac{1}{\sqrt{2}} ( | \text{acceleration 1} \rangle + | \text{acceleration 2} \rangle ) .
\end{align}
Due to the aforementioned correspondence between the Rindler and Schwarzschild spacetimes, we expect that the detector's response may provide insight into the quantum-gravitational effects produced by a mass-superposed Schwarzschild black hole.

Surprisingly, such a detector does not thermalize, not even at a temperature given by some function of the temperatures associates with the individual trajectories~\cite{foo2020unruhdewitt}. Even more surprisingly, the detector does not thermalize even if the two trajectories have the same acceleration and thus the same associated Unruh temperature but are simply translated in space in the direction of motion. This lack of thermalisation is due to quantum interference between the amplitudes associated with the different paths, which can be traced back to the fact that the detector interacts with two different sets of spatially translated field modes. 

We now move to the main focus of our framework and consider a detector {residing in a} superposition of spacetimes. 
Consider first a superposition of spacetime manifolds that are related by a coordinate transformation, e.g.\ spatial translation. This is  a scenario where ``two'' manifolds are related by a diffeomorphism and thus do not actually differ according to classical general relativity. Moreover, in any theory where the considered transformation is a symmetry of the dynamics,  any probability amplitude involving matter on such ``quantum'' spacetime is equivalent to an amplitude arising from a scenario in which the matter is in an appropriate superposition but on a single, classical spacetime manifold.

To understand this, note that the total Hilbert space of the relevant systems is $\mathcal{H} = \mathcal{H}_\text{UdW} \otimes \mathcal{H}_F \otimes \mathcal{H}_C$ where $\mathcal{H}_\text{UdW}$, $\mathcal{H}_F$ and $\mathcal{H}_C$ are respectively associated with the internal states of the detector, the field degrees of freedom (DoF), and a DoF which we call a control. Conceptually, the control is some ancillary system to which the field, spacetime and/or detector are coupled.

The interaction between all DoFs in our context very generally can be written as 
\begin{align}
    \hat{U} = \sum_i \hat{U}_i \otimes {_C |   i \rangle} \langle  i |_C 
\end{align}
where $\{|i\rangle_C\}$ are orthonormal states of the control (e.g.~position of the detector), and $\hat U_i $ is the interaction between the remaining DoFs for a given state of the control  (e.g.~interaction between the field and internal DoF of the detector for a given position of the latter). 
Given an initial superposition state of the control
\begin{align}
    | \psi \rangle_C &= \frac{1}{\sqrt{2}} ( |  1 \rangle_C + | 2 \rangle_C ) 
\end{align}
and an arbitrary state $|\varphi\rangle_{FD} $ of the   remaining DoFs (e.g.~the field and the internal DoF of an UdW detector) the conditional state of the latter two given the control is measured in  $|\psi\rangle_C$ reads
\begin{align}\label{eq:superpos_paths}
    {_C \langle \psi |} \hat{U} | \psi \rangle_C | \varphi\rangle_{FD} &= \frac{1}{2} ( \hat{U}_1 + \hat{U}_2 ) | \varphi_{FD} \rangle .
\end{align}
Note that the state of the control on which we condition is chosen to be the same as the initial state just for simplicity -- it can be arbitrary.

Suppose the control DoF describes the 
detector's position. The two control states can then be related using a possibly time-dependent function $L \equiv L ( \tau)$ relating the coordinates $\xi \equiv \xi(\tau)$ of the superposed  positions of the detector, such that  $|1\rangle_C\equiv|\xi\rangle_C$, $|2\rangle_C\equiv|\xi+L\rangle_C$. There thus exists an operator $\hat T_C$ implementing  spatial translations, such that $|\xi+L\rangle_C =\hat T_C(L)|\xi\rangle_C$. The corresponding change of coordinates  is implemented on the total system by an operator $\hat T_C(L)\otimes \hat{T}_F(L)$ and since it is a symmetry of the dynamics  we have~\cite{zych2018relativity}
\begin{align}\label{eq:equivalence}
\Big[ \hat U(\xi) + \hat U(\xi+L) \Big] |\varphi\rangle_{FD}  = { _C \Big\langle \xi \Big|} \Big[ \hat{U} + \hat{T}_F(L) \hat{U} \hat{T}^\dagger_F(L) \Big] \Big|\xi \Big\rangle_C \Big|\varphi \Big\rangle_{FD} .
\end{align}
The right-hand side describes the scenario where the detector is  traversing a single trajectory $\xi(\tau)$, but undergoes dynamics in which the field operators are enacted in a  ``superposition'' of translations. The left-hand side is Eq.~\eqref{eq:superpos_paths} for the special case of the control states being two positions $\xi, \xi+L$ of the detector, and describe a conditional state of the remaining DoFs given that the detector is prepared and measured in superposition (e.g.\ traversing a {Mach-Zender-type} interferometer). Equation~\eqref{eq:equivalence} is a formal statement of the fact that a scenario involving a particle in a spatially superposed spacetime is fully equivalent to a scenario where the particle follows superposed trajectories in one spacetime. The equivalence between the two can be interpreted as describing the same physical situation simply using two sets of coordinates related  by a ``superpostion'' of classical transformations \cite{zych2018relativity}, similar to scenarios arising in the context of quantum reference frames~\cite{giacomini2019}, i.e.~frames operationally established using quantum systems. This general equivalence further implies that a detector linearly coupled to the quantum field responds identically (i.e.\ has the same transition probability) in both scenarios. 

The crux of the argument is that only \textit{relative configurations} between the interacting systems are physically relevant, {for example  a superposition} of two distances between a pair of particles;  a global (joint) translation of all DoFs is irrelevant. 

\begin{figure}[h]
    \centering
    \includegraphics[width=0.35\linewidth]{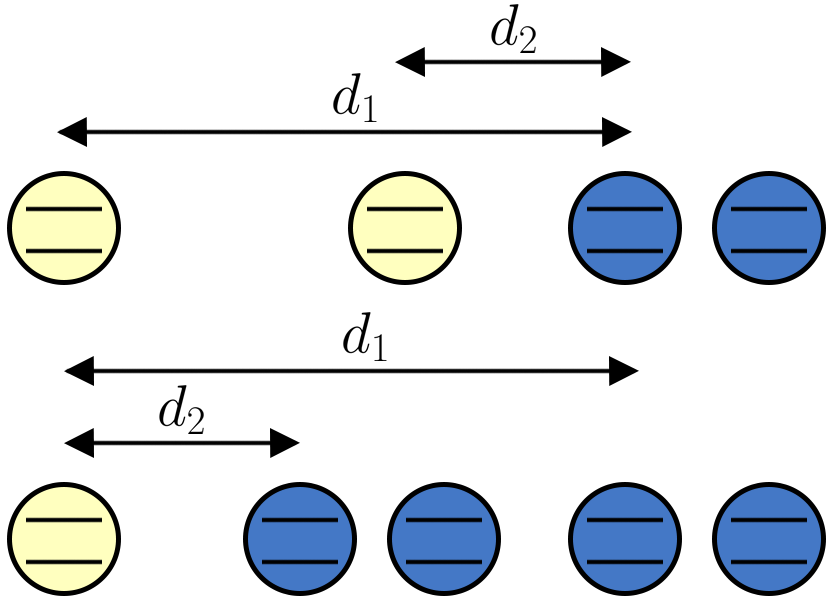}
    \caption{(Top) An illustration of the thought experiments of Bose \cite{bose2017spin} and Marletto \cite{marletto2017gravitationally}, where two spatially superposed source masses (depicted here as 2-level particles) interact via gravity. (Bottom) However such a system is equivalent to the case where one mass is on a fixed trajectory with the other undergoing two interference experiments in superposition. All relative distances remain unchanged. }
    \label{fig:marletto}
\end{figure}

This argument  also applies to the thought experiments by Bose \cite{bose2017spin} and Marletto  \cite{marletto2017gravitationally}, where two spatially superposed source masses gravitationally interact with each other, which is argued to lead to a superposition of spacetime metrics \cite{christodoulou2019possibility}. However, our argument shows that this is still equivalent to a situation in which one of the masses is on a fixed trajectory while the other undergoes two interference experiments in superposition -- closer and further from the source mass -- so that all relative distances are kept unchanged (see Fig.\ \ref{fig:marletto}). All the amplitudes (here leading to an entangled final state) are then also unchanged, but the explanation now only involves a superposition of distances between masses where the one designated as the source is fixed. Hence, one concludes that the spacetime is fixed as well. This highlights again the inherent ambiguity in the notion of ``spacetime superposition'' if the superposed amplitudes differ by a diffeomorphism (or any symmetry of the dynamics).

\begin{figure}[h]
    \centering
    \includegraphics[width=0.6\linewidth]{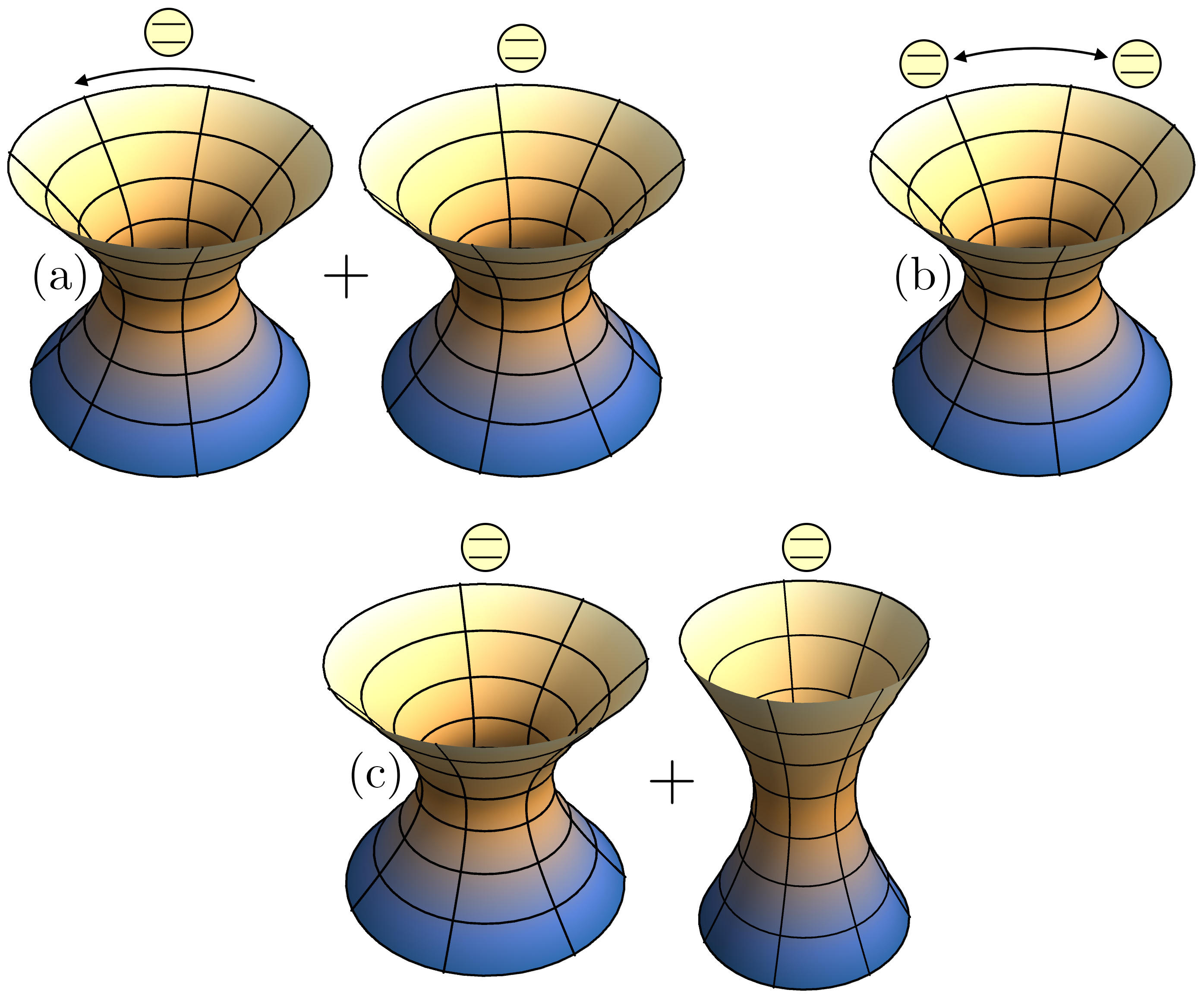}
    \caption{Visualisation of (a) a detector residing in a 
    de Sitter spacetime in a superposition of rotations, which is diffeomorphic to (b) a detector in a superposition of rotations on a classical de Sitter spacetime. Diagram  (c) is a visualization of de Sitter spacetime in a superposition of curvatures.  }
    \label{fig:visualisation}
\label{F1}    
\end{figure}

We can now apply this argument to the conceptually simple example of static de Sitter spacetime (and a basic model of our expanding universe), with line element
\begin{align}
    \D s^2 &= - \left( 1 - \frac{r^2}{l^2} \right) \D t^2 + \left( 1 - \frac{r^2}{l^2} \right)^{-1} \D r^2  + r^2 (\D\theta^2 + \sin^2\theta\D \phi^2)
\end{align}
where $l$ is the de Sitter length parametrizing the constant positive curvature of the spacetime, $1/l < r < \infty$, and $\theta, \phi \in [0,2\pi)$. Placing the detector in a superposition of spatially translated worldlines -- for example, a superposition of $r$'s or $\theta$'s -- is equivalent to one in which the detector resides on a manifold in a quantum superposition of spatial translations \cite{zych2018relativity, Foo:2020jmi}, {as illustrated in 
Figure~\ref{F1}(a,b).}

Compare this with a de Sitter universe whose curvature can be quantized and so is in a superposition of expansion rates (de Sitter lengths)~\cite{Foo:2020jmi}, see Figure~\ref{F1}(c). We thus associate the control DoF with the superposition 
\begin{align}
    | \psi \rangle_C &= \frac{1}{\sqrt{2}} ( | l_1 \rangle_C + |l_2 \rangle_C ). 
\end{align}
Here we have a truly novel situation: since the individual amplitudes of the superposition (each defined by different de Sitter length $l_1$ or $l_2$) are unique solutions to Einstein's field equations, there is no diffeomorphism that maps the spacetime superposition to a superposition of matter configurations  on a single metric having a fixed value of $l$. We might consider such a scenario to more uniquely describe a quantum-gravitational superposition, rather than a mere superposition of spatial configurations. 

We might also expect such a spacetime to elicit unambiguous signatures of a quantum-gravitational effects picked up by the detector. Interestingly however, owing to the equivalence between uniformly accelerated motion in flat Minkowski spacetime and static worldlines in expanding de Sitter spacetime, the effects produced by curvature-superposed de Sitter spacetime can be shown to be \textit{functionally} equivalent to those experienced by an acceleration-superposed detector in flat Minkowski spacetime
\cite{Foo:2020jmi}:
The detector's transition probability has a functional form that is
equivalent in both scenarios. This 
means that at least in this specific example, the UdW detector cannot unambiguously discern the effects produced by a genuine spacetime superposition of different curvatures, and that induced by a superposition of accelerated motions in flat spacetime. Note that this is not the same as mapping the two amplitudes arising from different curvatures to a single curved spacetime, but points out the equivalence of each amplitude to a flat-spacetime via the equivalence principle.

To determine whether matter coupled to spacetime via quantum fields is sensitive to genuinely quantum-gravitational effects, we can exploit a model system in which technical progress can be made: the $(2+1)$ Banados-Teitelboim-Zanelli (BTZ) dimensional black hole \cite{banadosPhysRevLett.69.1849}, obtained as a quotient of AdS-Rindler spacetime  \cite{hodgkinsonPhysRevD.86.064031,LANGLOIS20062027,Smith_2014}, giving the line element 
 \begin{align}
    \D s^2 &= - \left( \frac{r^2}{l^2} - M \right) \D t^2 + \left( \frac{r^2}{l^2} - M \right)^{-1} \D r^2 + r^2 \D \phi^2 ,
\end{align}
where $\sqrt{M}l < r < \infty$, $-\infty< t < \infty$, $\phi \in [0,2\pi)$ and $r_H = \sqrt{M}l$ is the radial coordinate of the event horizon. Our general approach can be now applied to the BTZ black hole in a superposition of masses $M_1$, $M_2$ \cite{Foo:2021fno}. We thus identify the control DoF with the mass of the black hole and obtain the conditional evolution of the remaining DoFs in the form of Eq.~\eqref{eq:superpos_paths} where $\hat 
U_i\equiv\hat U_{M_i}$ now describe the evolution of the field and detector for BTZ mass $M_i$. 

By tracing the field DoF we can even obtain the probability for the detector to register a field particle (Hawking quanta), which takes a general form
\begin{align}
    P_D &= \text{Tr}_F \left[ \frac{1}{2}(\hat U_{M_1} +\hat U_{M_1})| \varphi \rangle_{FD} \right] .
\end{align}

\begin{figure}[h]
    \centering
    \includegraphics[width=\linewidth]{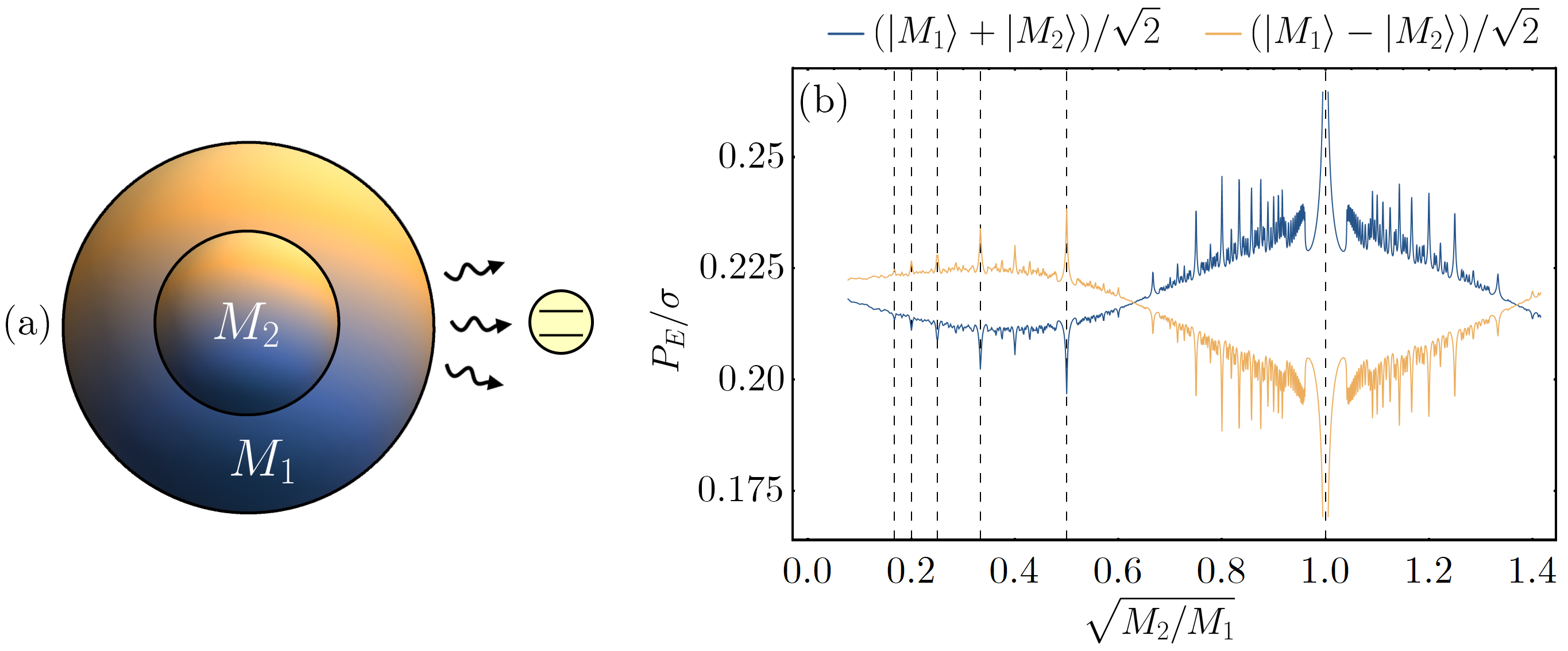}
    \caption{(a) Schematic diagram of the black hole mass superposition, with a detector situated at a fixed radial distance from the origin of coordinates. (b) Normalised probability $P_E/\sigma$ of the detector as a function of $\sqrt{M_2/M_1}$ ($\sigma$ is the timescale of the interaction). The measurement basis of the control is indicated in the legend. The dashed lines correspond to $\sqrt{M_2/M_1} = 1/n$ where $n = \{1,\hdots 6\}$. }
    \label{fig:plot1}
\end{figure}

Plotting the response of  the detector as a function of the mass ratio  $\sqrt{M_2/M_1}$ in Fig. \ref{fig:plot1}, we observe in both cases resonant peaks that occur at rational values of this ratio. 

We ascribe this behaviour to constructive interference  between the field modes associated with  topologically closed AdS spacetimes.  These ratios are commensurate with the quantum black hole conjecture of
Bekenstein \cite{bekensteinPhysRevD.7.2333,bekenstein2020quantum,reggePhysRev.108.1063}, who posited that a complete theory of quantum gravity must account for the treatment of black holes as quantum objects and that the horizon area (and hence its mass) is an adiabatic invariant with an associated discrete quantization \cite{hodPhysRevLett.81.4293}. Specifically, the allowed mass values of the BTZ black hole, assuming the Bohr-Sommerfeld quantization scheme for the horizon radius, are given by \cite{Kwon_2010}:
\begin{align}\label{eq19}
    r_H &= \sqrt{M}l = n , \qquad n = 1, 2, \hdots 
\end{align}
While our construction does not require that the superposed masses are quantized in integer values, we see that the detector responds uniquely to black hole mass superpositions with mass ratios corresponding to those predicted by Bekenstein's quantum gravity conjecture \cite{Foo:2021fno}. Our results demonstrate how a detector's transition probability can reveal genuinely quantum-gravitational properties of a spacetime superposition. 

Our approach has much potential
and scope for investigating spacetime superpositions and the quantum-gravitational effects they induce upon quantum matter 
Rather than developing a full-fledged quantization scheme for gravity itself, as numerous ``top-down'' approaches have attempted, we have shown that combining the simple but fundamental notion of quantum superposition with the equally foundational notion of spacetime leads to unique effects that can   in-principle be detected by a first-quantized system. Because we have approached the problem using the accessible tools of quantum field theory in curved spacetime, we are also able to ask and hopefully resolve further important questions of fundamental importance. 

One such question is how one may construct a ``quantum metric'', describing a superposed spacetime. Using our UdW detector model, one can in-principle measure the field correlations between different points in such a spacetime. Utilizing the fact that the strength of correlations in quantum field theory is an operational measure of spacetime distance \cite{Saravani_2016}, we can  write a  ``conditional quantum metric'' (i.e.\ conditioned on the measurement of the control system):
\begin{align}
    g_{\mu\nu} &\propto \lim_{x\to x' } \frac{\p}{\p x^\mu} \frac{\p}{\p x'^{\nu}} \sum_{i,j} \Big[  \underbrace{ {_F \langle \varphi |} \hat{\Phi}(x_i) \hat{\Phi}(x_j) | \varphi \rangle_F }_\text{two-point correlator}  \Big]^{\frac{2}{d-2}}
\end{align}
where $| \varphi\rangle_F$ is the state of the field, $\hat{\Phi}(x_i)$ is the field operator pulled back to the worldline $x_i$ of the $i$th amplitude of the spacetime superposition, and $d$ is the spacetime dimension. Calculating geodesics in such a spacetime poses an interesting question for further study. 

Another problem that has garnered recent interest has been the study of universal mechanisms of decoherence, particularly gravitational mechanisms. In our case, the interaction between the spacetime superposition, field, and matter is expected to decohere the system. However the extent of such decoherence remains to be seen. Conceivably one might be able to apply our approach to incorporate the ``self-decoherence'' of a black hole due to interaction with its own Hawking radiation, as suggested in refs.~\cite{Arrasmith_2019,Demers_1996}, and analyze the effects therein.

Finally, spacetime superpositions may find important applications in quantum foundations and quantum information. For example, they can be considered in the context of quantum-gravitational causal structures~\cite{zych2019} or quantum information processing in the presence of quantum gravity~\cite{Howl:2022oqz}. In particular, a detector situated outside a  mass-superposed black hole will experience a ``superposition'' of redshifts and proper times -- its interaction with the field being fundamentally indeterminate. This presents an important stepping stone towards understanding the notion of time in quantum gravity.

\bibliography{References.bib}

\begin{thebibliography}{49}%
\makeatletter
\providecommand \@ifxundefined [1]{%
 \@ifx{#1\undefined}
}%
\providecommand \@ifnum [1]{%
 \ifnum #1\expandafter \@firstoftwo
 \else \expandafter \@secondoftwo
 \fi
}%
\providecommand \@ifx [1]{%
 \ifx #1\expandafter \@firstoftwo
 \else \expandafter \@secondoftwo
 \fi
}%
\providecommand \natexlab [1]{#1}%
\providecommand \enquote  [1]{``#1''}%
\providecommand \bibnamefont  [1]{#1}%
\providecommand \bibfnamefont [1]{#1}%
\providecommand \citenamefont [1]{#1}%
\providecommand \href@noop [0]{\@secondoftwo}%
\providecommand \href [0]{\begingroup \@sanitize@url \@href}%
\providecommand \@href[1]{\@@startlink{#1}\@@href}%
\providecommand \@@href[1]{\endgroup#1\@@endlink}%
\providecommand \@sanitize@url [0]{\catcode `\\12\catcode `\$12\catcode
  `\&12\catcode `\#12\catcode `\^12\catcode `\_12\catcode `\%12\relax}%
\providecommand \@@startlink[1]{}%
\providecommand \@@endlink[0]{}%
\providecommand \url  [0]{\begingroup\@sanitize@url \@url }%
\providecommand \@url [1]{\endgroup\@href {#1}{\urlprefix }}%
\providecommand \urlprefix  [0]{URL }%
\providecommand \Eprint [0]{\href }%
\providecommand \doibase [0]{http://dx.doi.org/}%
\providecommand \selectlanguage [0]{\@gobble}%
\providecommand \bibinfo  [0]{\@secondoftwo}%
\providecommand \bibfield  [0]{\@secondoftwo}%
\providecommand \translation [1]{[#1]}%
\providecommand \BibitemOpen [0]{}%
\providecommand \bibitemStop [0]{}%
\providecommand \bibitemNoStop [0]{.\EOS\space}%
\providecommand \EOS [0]{\spacefactor3000\relax}%
\providecommand \BibitemShut  [1]{\csname bibitem#1\endcsname}%
\let\auto@bib@innerbib\@empty
\bibitem [{\citenamefont {Gubser}\ \emph {et~al.}(1998)\citenamefont {Gubser},
  \citenamefont {Klebanov},\ and\ \citenamefont {Polyakov}}]{gubser1998gauge}%
  \BibitemOpen
  \bibfield  {author} {\bibinfo {author} {\bibfnamefont {S.}~\bibnamefont
  {Gubser}}, \bibinfo {author} {\bibfnamefont {I.~R.}\ \bibnamefont
  {Klebanov}}, \ and\ \bibinfo {author} {\bibfnamefont {A.~M.}\ \bibnamefont
  {Polyakov}},\ }\href {\doibase 10.1016/S0370-2693(98)00377-3} {\bibfield
  {journal} {\bibinfo  {journal} {Phys. Lett. B}\ }\textbf {\bibinfo {volume}
  {428}},\ \bibinfo {pages} {105} (\bibinfo {year} {1998})},\ \Eprint
  {http://arxiv.org/abs/hep-th/9802109} {arXiv:hep-th/9802109} \BibitemShut
  {NoStop}%
\bibitem [{\citenamefont {Seiberg}\ and\ \citenamefont
  {Witten}(1999)}]{seiberg1999string}%
  \BibitemOpen
  \bibfield  {author} {\bibinfo {author} {\bibfnamefont {N.}~\bibnamefont
  {Seiberg}}\ and\ \bibinfo {author} {\bibfnamefont {E.}~\bibnamefont
  {Witten}},\ }\href {\doibase 10.1088/1126-6708/1999/09/032} {\bibfield
  {journal} {\bibinfo  {journal} {Journal of High Energy Physics}\ }\textbf
  {\bibinfo {volume} {1999}},\ \bibinfo {pages} {032} (\bibinfo {year}
  {1999})}\BibitemShut {NoStop}%
\bibitem [{\citenamefont {Polchinski}(1998)}]{polchinski1998string}%
  \BibitemOpen
  \bibfield  {author} {\bibinfo {author} {\bibfnamefont {J.}~\bibnamefont
  {Polchinski}},\ }\href@noop {} {\emph {\bibinfo {title} {String theory:
  Volume 2, superstring theory and beyond}}}\ (\bibinfo  {publisher} {Cambridge
  university press},\ \bibinfo {year} {1998})\BibitemShut {NoStop}%
\bibitem [{\citenamefont {Witten}(1995)}]{witten1995string}%
  \BibitemOpen
  \bibfield  {author} {\bibinfo {author} {\bibfnamefont {E.}~\bibnamefont
  {Witten}},\ }\href {\doibase 10.1016/0550-3213(95)00158-O} {\bibfield
  {journal} {\bibinfo  {journal} {Nucl. Phys. B}\ }\textbf {\bibinfo {volume}
  {443}},\ \bibinfo {pages} {85} (\bibinfo {year} {1995})}\BibitemShut
  {NoStop}%
\bibitem [{\citenamefont {Rovelli}(1998)}]{rovelli2008loop}%
  \BibitemOpen
  \bibfield  {author} {\bibinfo {author} {\bibfnamefont {C.}~\bibnamefont
  {Rovelli}},\ }\href {\doibase 10.12942/lrr-1998-1} {\bibfield  {journal}
  {\bibinfo  {journal} {Living Rev. Rel.}\ }\textbf {\bibinfo {volume} {1}},\
  \bibinfo {pages} {1} (\bibinfo {year} {1998})}\BibitemShut {NoStop}%
\bibitem [{\citenamefont {Rovelli}\ and\ \citenamefont
  {Vidotto}(2014)}]{rovelli2014covariant}%
  \BibitemOpen
  \bibfield  {author} {\bibinfo {author} {\bibfnamefont {C.}~\bibnamefont
  {Rovelli}}\ and\ \bibinfo {author} {\bibfnamefont {F.}~\bibnamefont
  {Vidotto}},\ }\href@noop {} {\emph {\bibinfo {title} {Covariant loop quantum
  gravity: an elementary introduction to quantum gravity and spinfoam
  theory}}}\ (\bibinfo  {publisher} {Cambridge University Press},\ \bibinfo
  {year} {2014})\BibitemShut {NoStop}%
\bibitem [{\citenamefont {Thiemann}(2003)}]{thiemann2003lectures}%
  \BibitemOpen
  \bibfield  {author} {\bibinfo {author} {\bibfnamefont {T.}~\bibnamefont
  {Thiemann}},\ }in\ \href@noop {} {\emph {\bibinfo {booktitle} {Quantum
  gravity}}}\ (\bibinfo  {publisher} {Springer},\ \bibinfo {year} {2003})\ pp.\
  \bibinfo {pages} {41--135}\BibitemShut {NoStop}%
\bibitem [{\citenamefont {Dyson}(2013)}]{dyson2014graviton}%
  \BibitemOpen
  \bibfield  {author} {\bibinfo {author} {\bibfnamefont {F.}~\bibnamefont
  {Dyson}},\ }\href {\doibase 10.1142/S0217751X1330041X} {\bibfield  {journal}
  {\bibinfo  {journal} {Int. J. Mod. Phys. A}\ }\textbf {\bibinfo {volume}
  {28}},\ \bibinfo {pages} {1330041} (\bibinfo {year} {2013})}\BibitemShut
  {NoStop}%
\bibitem [{\citenamefont {Garay}(1995)}]{garay1995quantum}%
  \BibitemOpen
  \bibfield  {author} {\bibinfo {author} {\bibfnamefont {L.~J.}\ \bibnamefont
  {Garay}},\ }\href {\doibase 10.1142/S0217751X95000085} {\bibfield  {journal}
  {\bibinfo  {journal} {Int. J. Mod. Phys. A}\ }\textbf {\bibinfo {volume}
  {10}},\ \bibinfo {pages} {145} (\bibinfo {year} {1995})}\BibitemShut
  {NoStop}%
\bibitem [{\citenamefont {Carlip}(2008)}]{carlip2008quantum}%
  \BibitemOpen
  \bibfield  {author} {\bibinfo {author} {\bibfnamefont {S.}~\bibnamefont
  {Carlip}},\ }\href {\doibase 10.1088/0264-9381/25/15/154010} {\bibfield
  {journal} {\bibinfo  {journal} {Classical and Quantum Gravity}\ }\textbf
  {\bibinfo {volume} {25}},\ \bibinfo {pages} {154010} (\bibinfo {year}
  {2008})}\BibitemShut {NoStop}%
\bibitem [{\citenamefont {Bose}\ \emph {et~al.}(2017)\citenamefont {Bose},
  \citenamefont {Mazumdar}, \citenamefont {Morley}, \citenamefont {Ulbricht},
  \citenamefont {Toro\ifmmode~\check{s}\else \v{s}\fi{}}, \citenamefont
  {Paternostro}, \citenamefont {Geraci}, \citenamefont {Barker}, \citenamefont
  {Kim},\ and\ \citenamefont {Milburn}}]{bose2017spin}%
  \BibitemOpen
  \bibfield  {author} {\bibinfo {author} {\bibfnamefont {S.}~\bibnamefont
  {Bose}}, \bibinfo {author} {\bibfnamefont {A.}~\bibnamefont {Mazumdar}},
  \bibinfo {author} {\bibfnamefont {G.~W.}\ \bibnamefont {Morley}}, \bibinfo
  {author} {\bibfnamefont {H.}~\bibnamefont {Ulbricht}}, \bibinfo {author}
  {\bibfnamefont {M.}~\bibnamefont {Toro\ifmmode~\check{s}\else \v{s}\fi{}}},
  \bibinfo {author} {\bibfnamefont {M.}~\bibnamefont {Paternostro}}, \bibinfo
  {author} {\bibfnamefont {A.~A.}\ \bibnamefont {Geraci}}, \bibinfo {author}
  {\bibfnamefont {P.~F.}\ \bibnamefont {Barker}}, \bibinfo {author}
  {\bibfnamefont {M.~S.}\ \bibnamefont {Kim}}, \ and\ \bibinfo {author}
  {\bibfnamefont {G.}~\bibnamefont {Milburn}},\ }\href {\doibase
  10.1103/PhysRevLett.119.240401} {\bibfield  {journal} {\bibinfo  {journal}
  {Phys. Rev. Lett.}\ }\textbf {\bibinfo {volume} {119}},\ \bibinfo {pages}
  {240401} (\bibinfo {year} {2017})}\BibitemShut {NoStop}%
\bibitem [{\citenamefont {Marletto}\ and\ \citenamefont
  {Vedral}(2017)}]{marletto2017gravitationally}%
  \BibitemOpen
  \bibfield  {author} {\bibinfo {author} {\bibfnamefont {C.}~\bibnamefont
  {Marletto}}\ and\ \bibinfo {author} {\bibfnamefont {V.}~\bibnamefont
  {Vedral}},\ }\href {\doibase 10.1103/PhysRevLett.119.240402} {\bibfield
  {journal} {\bibinfo  {journal} {Phys. Rev. Lett.}\ }\textbf {\bibinfo
  {volume} {119}},\ \bibinfo {pages} {240402} (\bibinfo {year}
  {2017})}\BibitemShut {NoStop}%
\bibitem [{\citenamefont {Christodoulou}\ and\ \citenamefont
  {Rovelli}(2019)}]{christodoulou2019possibility}%
  \BibitemOpen
  \bibfield  {author} {\bibinfo {author} {\bibfnamefont {M.}~\bibnamefont
  {Christodoulou}}\ and\ \bibinfo {author} {\bibfnamefont {C.}~\bibnamefont
  {Rovelli}},\ }\href {\doibase 10.1016/j.physletb.2019.03.015} {\bibfield
  {journal} {\bibinfo  {journal} {Physics Letters B}\ }\textbf {\bibinfo
  {volume} {792}},\ \bibinfo {pages} {64–68} (\bibinfo {year}
  {2019})}\BibitemShut {NoStop}%
\bibitem [{\citenamefont {Arrasmith}\ \emph {et~al.}(2019)\citenamefont
  {Arrasmith}, \citenamefont {Albrecht},\ and\ \citenamefont
  {Zurek}}]{Arrasmith_2019}%
  \BibitemOpen
  \bibfield  {author} {\bibinfo {author} {\bibfnamefont {A.}~\bibnamefont
  {Arrasmith}}, \bibinfo {author} {\bibfnamefont {A.}~\bibnamefont {Albrecht}},
  \ and\ \bibinfo {author} {\bibfnamefont {W.~H.}\ \bibnamefont {Zurek}},\
  }\href {\doibase 10.1038/s41467-019-08426-4} {\bibfield  {journal} {\bibinfo
  {journal} {Nature Communications}\ }\textbf {\bibinfo {volume} {10}}
  (\bibinfo {year} {2019}),\ 10.1038/s41467-019-08426-4}\BibitemShut {NoStop}%
\bibitem [{\citenamefont {Demers}\ and\ \citenamefont
  {Kiefer}(1996)}]{Demers_1996}%
  \BibitemOpen
  \bibfield  {author} {\bibinfo {author} {\bibfnamefont {J.-G.}\ \bibnamefont
  {Demers}}\ and\ \bibinfo {author} {\bibfnamefont {C.}~\bibnamefont
  {Kiefer}},\ }\href {\doibase 10.1103/physrevd.53.7050} {\bibfield  {journal}
  {\bibinfo  {journal} {Physical Review D}\ }\textbf {\bibinfo {volume} {53}},\
  \bibinfo {pages} {7050} (\bibinfo {year} {1996})}\BibitemShut {NoStop}%
\bibitem [{\citenamefont {Allali}\ and\ \citenamefont
  {Hertzberg}(2021)}]{allaliPhysRevLett.127.031301}%
  \BibitemOpen
  \bibfield  {author} {\bibinfo {author} {\bibfnamefont {I.~J.}\ \bibnamefont
  {Allali}}\ and\ \bibinfo {author} {\bibfnamefont {M.~P.}\ \bibnamefont
  {Hertzberg}},\ }\href {\doibase 10.1103/PhysRevLett.127.031301} {\bibfield
  {journal} {\bibinfo  {journal} {Phys. Rev. Lett.}\ }\textbf {\bibinfo
  {volume} {127}},\ \bibinfo {pages} {031301} (\bibinfo {year}
  {2021})}\BibitemShut {NoStop}%
\bibitem [{\citenamefont {Schut}\ \emph {et~al.}(2022)\citenamefont {Schut},
  \citenamefont {Tilly}, \citenamefont {Marshman}, \citenamefont {Bose},\ and\
  \citenamefont {Mazumdar}}]{bosePhysRevA.105.032411}%
  \BibitemOpen
  \bibfield  {author} {\bibinfo {author} {\bibfnamefont {M.}~\bibnamefont
  {Schut}}, \bibinfo {author} {\bibfnamefont {J.}~\bibnamefont {Tilly}},
  \bibinfo {author} {\bibfnamefont {R.~J.}\ \bibnamefont {Marshman}}, \bibinfo
  {author} {\bibfnamefont {S.}~\bibnamefont {Bose}}, \ and\ \bibinfo {author}
  {\bibfnamefont {A.}~\bibnamefont {Mazumdar}},\ }\href {\doibase
  10.1103/PhysRevA.105.032411} {\bibfield  {journal} {\bibinfo  {journal}
  {Phys. Rev. A}\ }\textbf {\bibinfo {volume} {105}},\ \bibinfo {pages}
  {032411} (\bibinfo {year} {2022})}\BibitemShut {NoStop}%
\bibitem [{\citenamefont {Howl}\ \emph {et~al.}(2021)\citenamefont {Howl},
  \citenamefont {Vedral}, \citenamefont {Naik}, \citenamefont {Christodoulou},
  \citenamefont {Rovelli},\ and\ \citenamefont
  {Iyer}}]{howlPRXQuantum.2.010325}%
  \BibitemOpen
  \bibfield  {author} {\bibinfo {author} {\bibfnamefont {R.}~\bibnamefont
  {Howl}}, \bibinfo {author} {\bibfnamefont {V.}~\bibnamefont {Vedral}},
  \bibinfo {author} {\bibfnamefont {D.}~\bibnamefont {Naik}}, \bibinfo {author}
  {\bibfnamefont {M.}~\bibnamefont {Christodoulou}}, \bibinfo {author}
  {\bibfnamefont {C.}~\bibnamefont {Rovelli}}, \ and\ \bibinfo {author}
  {\bibfnamefont {A.}~\bibnamefont {Iyer}},\ }\href {\doibase
  10.1103/PRXQuantum.2.010325} {\bibfield  {journal} {\bibinfo  {journal} {PRX
  Quantum}\ }\textbf {\bibinfo {volume} {2}},\ \bibinfo {pages} {010325}
  (\bibinfo {year} {2021})}\BibitemShut {NoStop}%
\bibitem [{\citenamefont {Carney}\ \emph {et~al.}(2021)\citenamefont {Carney},
  \citenamefont {M\"uller},\ and\ \citenamefont
  {Taylor}}]{carneyPRXQuantum.2.030330}%
  \BibitemOpen
  \bibfield  {author} {\bibinfo {author} {\bibfnamefont {D.}~\bibnamefont
  {Carney}}, \bibinfo {author} {\bibfnamefont {H.}~\bibnamefont {M\"uller}}, \
  and\ \bibinfo {author} {\bibfnamefont {J.~M.}\ \bibnamefont {Taylor}},\
  }\href {\doibase 10.1103/PRXQuantum.2.030330} {\bibfield  {journal} {\bibinfo
   {journal} {PRX Quantum}\ }\textbf {\bibinfo {volume} {2}},\ \bibinfo {pages}
  {030330} (\bibinfo {year} {2021})}\BibitemShut {NoStop}%
\bibitem [{\citenamefont {Nova}\ \emph {et~al.}(2019)\citenamefont {Nova},
  \citenamefont {Golubkov}, \citenamefont {Kolobov},\ and\ \citenamefont
  {Steinhauer}}]{Steihnauer2019Nature}%
  \BibitemOpen
  \bibfield  {author} {\bibinfo {author} {\bibfnamefont {J.}~\bibnamefont
  {Nova}}, \bibinfo {author} {\bibfnamefont {K.}~\bibnamefont {Golubkov}},
  \bibinfo {author} {\bibfnamefont {V.}~\bibnamefont {Kolobov}}, \ and\
  \bibinfo {author} {\bibfnamefont {J.}~\bibnamefont {Steinhauer}},\ }\href
  {\doibase 10.1038/s41586-019-1241-0} {\bibfield  {journal} {\bibinfo
  {journal} {Nature}\ }\textbf {\bibinfo {volume} {569}},\ \bibinfo {pages}
  {688} (\bibinfo {year} {2019})}\BibitemShut {NoStop}%
\bibitem [{\citenamefont {Barceló}\ \emph {et~al.}(2021)\citenamefont
  {Barceló}, \citenamefont {Garay},\ and\ \citenamefont
  {García-Moreno}}]{barcelo2021superposing}%
  \BibitemOpen
  \bibfield  {author} {\bibinfo {author} {\bibfnamefont {C.}~\bibnamefont
  {Barceló}}, \bibinfo {author} {\bibfnamefont {L.~J.}\ \bibnamefont {Garay}},
  \ and\ \bibinfo {author} {\bibfnamefont {G.}~\bibnamefont {García-Moreno}},\
  }\href@noop {} {\enquote {\bibinfo {title} {Superposing spacetimes: lessons
  from analogue gravity},}\ } (\bibinfo {year} {2021}),\ \Eprint
  {http://arxiv.org/abs/2104.15078} {arXiv:2104.15078 [gr-qc]} \BibitemShut
  {NoStop}%
\bibitem [{\citenamefont {Zych}\ \emph {et~al.}(2018)\citenamefont {Zych},
  \citenamefont {Costa},\ and\ \citenamefont {Ralph}}]{zych2018relativity}%
  \BibitemOpen
  \bibfield  {author} {\bibinfo {author} {\bibfnamefont {M.}~\bibnamefont
  {Zych}}, \bibinfo {author} {\bibfnamefont {F.}~\bibnamefont {Costa}}, \ and\
  \bibinfo {author} {\bibfnamefont {T.~C.}\ \bibnamefont {Ralph}},\ }\href@noop
  {} {\enquote {\bibinfo {title} {Relativity of quantum superpositions},}\ }
  (\bibinfo {year} {2018}),\ \Eprint {http://arxiv.org/abs/1809.04999}
  {arXiv:1809.04999 [quant-ph]} \BibitemShut {NoStop}%
\bibitem [{\citenamefont {Foo}\ \emph {et~al.}(2021{\natexlab{a}})\citenamefont
  {Foo}, \citenamefont {Mann},\ and\ \citenamefont {Zych}}]{Foo_2021}%
  \BibitemOpen
  \bibfield  {author} {\bibinfo {author} {\bibfnamefont {J.}~\bibnamefont
  {Foo}}, \bibinfo {author} {\bibfnamefont {R.~B.}\ \bibnamefont {Mann}}, \
  and\ \bibinfo {author} {\bibfnamefont {M.}~\bibnamefont {Zych}},\ }\href
  {\doibase 10.1088/1361-6382/abf1c4} {\bibfield  {journal} {\bibinfo
  {journal} {Classical and Quantum Gravity}\ }\textbf {\bibinfo {volume}
  {38}},\ \bibinfo {pages} {115010} (\bibinfo {year}
  {2021}{\natexlab{a}})}\BibitemShut {NoStop}%
\bibitem [{\citenamefont {Unruh}\ and\ \citenamefont
  {Wald}(1984)}]{unruh1984happens}%
  \BibitemOpen
  \bibfield  {author} {\bibinfo {author} {\bibfnamefont {W.~G.}\ \bibnamefont
  {Unruh}}\ and\ \bibinfo {author} {\bibfnamefont {R.~M.}\ \bibnamefont
  {Wald}},\ }\href {\doibase 10.1103/PhysRevD.29.1047} {\bibfield  {journal}
  {\bibinfo  {journal} {Phys. Rev. D}\ }\textbf {\bibinfo {volume} {29}},\
  \bibinfo {pages} {1047} (\bibinfo {year} {1984})}\BibitemShut {NoStop}%
\bibitem [{\citenamefont {Ng}\ \emph {et~al.}(2014)\citenamefont {Ng},
  \citenamefont {Hodgkinson}, \citenamefont {Louko}, \citenamefont {Mann},\
  and\ \citenamefont {Martin-Martinez}}]{ng2014unruh}%
  \BibitemOpen
  \bibfield  {author} {\bibinfo {author} {\bibfnamefont {K.~K.}\ \bibnamefont
  {Ng}}, \bibinfo {author} {\bibfnamefont {L.}~\bibnamefont {Hodgkinson}},
  \bibinfo {author} {\bibfnamefont {J.}~\bibnamefont {Louko}}, \bibinfo
  {author} {\bibfnamefont {R.~B.}\ \bibnamefont {Mann}}, \ and\ \bibinfo
  {author} {\bibfnamefont {E.}~\bibnamefont {Martin-Martinez}},\ }\href
  {\doibase 10.1103/PhysRevD.90.064003} {\bibfield  {journal} {\bibinfo
  {journal} {Phys. Rev. D}\ }\textbf {\bibinfo {volume} {90}},\ \bibinfo
  {pages} {064003} (\bibinfo {year} {2014})}\BibitemShut {NoStop}%
\bibitem [{\citenamefont {Ng}\ \emph {et~al.}(2017)\citenamefont {Ng},
  \citenamefont {Mann},\ and\ \citenamefont
  {Mart\'{\i}n-Mart\'{\i}nez}}]{ng2017over}%
  \BibitemOpen
  \bibfield  {author} {\bibinfo {author} {\bibfnamefont {K.~K.}\ \bibnamefont
  {Ng}}, \bibinfo {author} {\bibfnamefont {R.~B.}\ \bibnamefont {Mann}}, \ and\
  \bibinfo {author} {\bibfnamefont {E.}~\bibnamefont
  {Mart\'{\i}n-Mart\'{\i}nez}},\ }\href {\doibase 10.1103/PhysRevD.96.085004}
  {\bibfield  {journal} {\bibinfo  {journal} {Phys. Rev. D}\ }\textbf {\bibinfo
  {volume} {96}},\ \bibinfo {pages} {085004} (\bibinfo {year}
  {2017})}\BibitemShut {NoStop}%
\bibitem [{\citenamefont {Ng}\ \emph {et~al.}(2018)\citenamefont {Ng},
  \citenamefont {Mann},\ and\ \citenamefont
  {Mart\'{\i}n-Mart\'{\i}nez}}]{ng2018new}%
  \BibitemOpen
  \bibfield  {author} {\bibinfo {author} {\bibfnamefont {K.~K.}\ \bibnamefont
  {Ng}}, \bibinfo {author} {\bibfnamefont {R.~B.}\ \bibnamefont {Mann}}, \ and\
  \bibinfo {author} {\bibfnamefont {E.}~\bibnamefont
  {Mart\'{\i}n-Mart\'{\i}nez}},\ }\href {\doibase 10.1103/PhysRevD.97.125011}
  {\bibfield  {journal} {\bibinfo  {journal} {Phys. Rev. D}\ }\textbf {\bibinfo
  {volume} {97}},\ \bibinfo {pages} {125011} (\bibinfo {year}
  {2018})}\BibitemShut {NoStop}%
\bibitem [{\citenamefont {Henderson}\ \emph
  {et~al.}(2020{\natexlab{a}})\citenamefont {Henderson}, \citenamefont
  {Hennigar}, \citenamefont {Mann}, \citenamefont {Smith},\ and\ \citenamefont
  {Zhang}}]{henderson2019btz}%
  \BibitemOpen
  \bibfield  {author} {\bibinfo {author} {\bibfnamefont {L.~J.}\ \bibnamefont
  {Henderson}}, \bibinfo {author} {\bibfnamefont {R.~A.}\ \bibnamefont
  {Hennigar}}, \bibinfo {author} {\bibfnamefont {R.~B.}\ \bibnamefont {Mann}},
  \bibinfo {author} {\bibfnamefont {A.~R.}\ \bibnamefont {Smith}}, \ and\
  \bibinfo {author} {\bibfnamefont {J.}~\bibnamefont {Zhang}},\ }\href
  {\doibase 10.1016/j.physletb.2020.135732} {\bibfield  {journal} {\bibinfo
  {journal} {Phys. Lett. B}\ }\textbf {\bibinfo {volume} {809}},\ \bibinfo
  {pages} {135732} (\bibinfo {year} {2020}{\natexlab{a}})}\BibitemShut
  {NoStop}%
\bibitem [{\citenamefont {Hodgkinson}\ and\ \citenamefont
  {Louko}(2012{\natexlab{a}})}]{hodgkinson2012static}%
  \BibitemOpen
  \bibfield  {author} {\bibinfo {author} {\bibfnamefont {L.}~\bibnamefont
  {Hodgkinson}}\ and\ \bibinfo {author} {\bibfnamefont {J.}~\bibnamefont
  {Louko}},\ }\href@noop {} {\bibfield  {journal} {\bibinfo  {journal}
  {Physical Review D}\ }\textbf {\bibinfo {volume} {86}},\ \bibinfo {pages}
  {064031} (\bibinfo {year} {2012}{\natexlab{a}})}\BibitemShut {NoStop}%
\bibitem [{\citenamefont {Hodgkinson}\ and\ \citenamefont
  {Louko}(2012{\natexlab{b}})}]{hodgkinson2012often}%
  \BibitemOpen
  \bibfield  {author} {\bibinfo {author} {\bibfnamefont {L.}~\bibnamefont
  {Hodgkinson}}\ and\ \bibinfo {author} {\bibfnamefont {J.}~\bibnamefont
  {Louko}},\ }\href {\doibase 10.1063/1.4739453} {\bibfield  {journal}
  {\bibinfo  {journal} {Journal of Mathematical Physics}\ }\textbf {\bibinfo
  {volume} {53}},\ \bibinfo {pages} {082301} (\bibinfo {year}
  {2012}{\natexlab{b}})},\ \Eprint
  {http://arxiv.org/abs/https://doi.org/10.1063/1.4739453}
  {https://doi.org/10.1063/1.4739453} \BibitemShut {NoStop}%
\bibitem [{\citenamefont {Henderson}\ \emph
  {et~al.}(2020{\natexlab{b}})\citenamefont {Henderson}, \citenamefont
  {Belenchia}, \citenamefont {Castro-Ruiz}, \citenamefont {Budroni},
  \citenamefont {Zych}, \citenamefont {Časlav Brukner},\ and\ \citenamefont
  {Mann}}]{henderson2020quantum}%
  \BibitemOpen
  \bibfield  {author} {\bibinfo {author} {\bibfnamefont {L.~J.}\ \bibnamefont
  {Henderson}}, \bibinfo {author} {\bibfnamefont {A.}~\bibnamefont
  {Belenchia}}, \bibinfo {author} {\bibfnamefont {E.}~\bibnamefont
  {Castro-Ruiz}}, \bibinfo {author} {\bibfnamefont {C.}~\bibnamefont
  {Budroni}}, \bibinfo {author} {\bibfnamefont {M.}~\bibnamefont {Zych}},
  \bibinfo {author} {\bibnamefont {Časlav Brukner}}, \ and\ \bibinfo {author}
  {\bibfnamefont {R.~B.}\ \bibnamefont {Mann}},\ }\href@noop {} {\enquote
  {\bibinfo {title} {Quantum temporal superposition: the case of qft},}\ }
  (\bibinfo {year} {2020}{\natexlab{b}}),\ \Eprint
  {http://arxiv.org/abs/2002.06208} {arXiv:2002.06208 [quant-ph]} \BibitemShut
  {NoStop}%
\bibitem [{\citenamefont {Foo}\ \emph {et~al.}(2020)\citenamefont {Foo},
  \citenamefont {Onoe},\ and\ \citenamefont {Zych}}]{foo2020unruhdewitt}%
  \BibitemOpen
  \bibfield  {author} {\bibinfo {author} {\bibfnamefont {J.}~\bibnamefont
  {Foo}}, \bibinfo {author} {\bibfnamefont {S.}~\bibnamefont {Onoe}}, \ and\
  \bibinfo {author} {\bibfnamefont {M.}~\bibnamefont {Zych}},\ }\href {\doibase
  10.1103/PhysRevD.102.085013} {\bibfield  {journal} {\bibinfo  {journal}
  {Phys. Rev. D}\ }\textbf {\bibinfo {volume} {102}},\ \bibinfo {pages}
  {085013} (\bibinfo {year} {2020})}\BibitemShut {NoStop}%
\bibitem [{\citenamefont {Barbado}\ \emph {et~al.}(2020)\citenamefont
  {Barbado}, \citenamefont {Castro-Ruiz}, \citenamefont {Apadula},\ and\
  \citenamefont {Brukner}}]{barbado2020unruh}%
  \BibitemOpen
  \bibfield  {author} {\bibinfo {author} {\bibfnamefont {L.~C.}\ \bibnamefont
  {Barbado}}, \bibinfo {author} {\bibfnamefont {E.}~\bibnamefont
  {Castro-Ruiz}}, \bibinfo {author} {\bibfnamefont {L.}~\bibnamefont
  {Apadula}}, \ and\ \bibinfo {author} {\bibfnamefont {C.}~\bibnamefont
  {Brukner}},\ }\href {\doibase 10.1103/PhysRevD.102.045002} {\bibfield
  {journal} {\bibinfo  {journal} {Phys. Rev. D}\ }\textbf {\bibinfo {volume}
  {102}},\ \bibinfo {pages} {045002} (\bibinfo {year} {2020})}\BibitemShut
  {NoStop}%
\bibitem [{\citenamefont {Foo}\ \emph {et~al.}(2021{\natexlab{b}})\citenamefont
  {Foo}, \citenamefont {Onoe}, \citenamefont {Mann},\ and\ \citenamefont
  {Zych}}]{foo2020thermality}%
  \BibitemOpen
  \bibfield  {author} {\bibinfo {author} {\bibfnamefont {J.}~\bibnamefont
  {Foo}}, \bibinfo {author} {\bibfnamefont {S.}~\bibnamefont {Onoe}}, \bibinfo
  {author} {\bibfnamefont {R.~B.}\ \bibnamefont {Mann}}, \ and\ \bibinfo
  {author} {\bibfnamefont {M.}~\bibnamefont {Zych}},\ }\href {\doibase
  10.1103/PhysRevResearch.3.043056} {\bibfield  {journal} {\bibinfo  {journal}
  {Phys. Rev. Research}\ }\textbf {\bibinfo {volume} {3}},\ \bibinfo {pages}
  {043056} (\bibinfo {year} {2021}{\natexlab{b}})}\BibitemShut {NoStop}%
\bibitem [{\citenamefont {Giacomini}\ \emph {et~al.}(2019)\citenamefont
  {Giacomini}, \citenamefont {Castro-Ruiz},\ and\ \citenamefont
  {Brukner}}]{giacomini2019}%
  \BibitemOpen
  \bibfield  {author} {\bibinfo {author} {\bibfnamefont {F.}~\bibnamefont
  {Giacomini}}, \bibinfo {author} {\bibfnamefont {E.}~\bibnamefont
  {Castro-Ruiz}}, \ and\ \bibinfo {author} {\bibfnamefont {C.}~\bibnamefont
  {Brukner}},\ }\href {\doibase https://doi.org/10.1038/s41467-018-08155-0}
  {\bibfield  {journal} {\bibinfo  {journal} {Nat Commun}\ }\textbf {\bibinfo
  {volume} {10}} (\bibinfo {year} {2019}),\
  https://doi.org/10.1038/s41467-018-08155-0}\BibitemShut {NoStop}%
\bibitem [{\citenamefont {Foo}\ \emph {et~al.}(2021{\natexlab{c}})\citenamefont
  {Foo}, \citenamefont {Mann},\ and\ \citenamefont {Zych}}]{Foo:2020jmi}%
  \BibitemOpen
  \bibfield  {author} {\bibinfo {author} {\bibfnamefont {J.}~\bibnamefont
  {Foo}}, \bibinfo {author} {\bibfnamefont {R.~B.}\ \bibnamefont {Mann}}, \
  and\ \bibinfo {author} {\bibfnamefont {M.}~\bibnamefont {Zych}},\ }\href
  {\doibase 10.1088/1361-6382/abf1c4} {\bibfield  {journal} {\bibinfo
  {journal} {Class. Quant. Grav.}\ }\textbf {\bibinfo {volume} {38}},\ \bibinfo
  {pages} {115010} (\bibinfo {year} {2021}{\natexlab{c}})},\ \Eprint
  {http://arxiv.org/abs/2012.10025} {arXiv:2012.10025 [gr-qc]} \BibitemShut
  {NoStop}%
\bibitem [{\citenamefont {Ba\~nados}\ \emph {et~al.}(1992)\citenamefont
  {Ba\~nados}, \citenamefont {Teitelboim},\ and\ \citenamefont
  {Zanelli}}]{banadosPhysRevLett.69.1849}%
  \BibitemOpen
  \bibfield  {author} {\bibinfo {author} {\bibfnamefont {M.}~\bibnamefont
  {Ba\~nados}}, \bibinfo {author} {\bibfnamefont {C.}~\bibnamefont
  {Teitelboim}}, \ and\ \bibinfo {author} {\bibfnamefont {J.}~\bibnamefont
  {Zanelli}},\ }\href {\doibase 10.1103/PhysRevLett.69.1849} {\bibfield
  {journal} {\bibinfo  {journal} {Phys. Rev. Lett.}\ }\textbf {\bibinfo
  {volume} {69}},\ \bibinfo {pages} {1849} (\bibinfo {year}
  {1992})}\BibitemShut {NoStop}%
\bibitem [{\citenamefont {Hodgkinson}\ and\ \citenamefont
  {Louko}(2012{\natexlab{c}})}]{hodgkinsonPhysRevD.86.064031}%
  \BibitemOpen
  \bibfield  {author} {\bibinfo {author} {\bibfnamefont {L.}~\bibnamefont
  {Hodgkinson}}\ and\ \bibinfo {author} {\bibfnamefont {J.}~\bibnamefont
  {Louko}},\ }\href {\doibase 10.1103/PhysRevD.86.064031} {\bibfield  {journal}
  {\bibinfo  {journal} {Phys. Rev. D}\ }\textbf {\bibinfo {volume} {86}},\
  \bibinfo {pages} {064031} (\bibinfo {year} {2012}{\natexlab{c}})}\BibitemShut
  {NoStop}%
\bibitem [{\citenamefont {Langlois}(2006)}]{LANGLOIS20062027}%
  \BibitemOpen
  \bibfield  {author} {\bibinfo {author} {\bibfnamefont {P.}~\bibnamefont
  {Langlois}},\ }\href {\doibase https://doi.org/10.1016/j.aop.2006.01.013}
  {\bibfield  {journal} {\bibinfo  {journal} {Annals of Physics}\ }\textbf
  {\bibinfo {volume} {321}},\ \bibinfo {pages} {2027} (\bibinfo {year}
  {2006})}\BibitemShut {NoStop}%
\bibitem [{\citenamefont {Smith}\ and\ \citenamefont
  {Mann}(2014)}]{Smith_2014}%
  \BibitemOpen
  \bibfield  {author} {\bibinfo {author} {\bibfnamefont {A.~R.~H.}\
  \bibnamefont {Smith}}\ and\ \bibinfo {author} {\bibfnamefont {R.~B.}\
  \bibnamefont {Mann}},\ }\href {\doibase 10.1088/0264-9381/31/8/082001}
  {\bibfield  {journal} {\bibinfo  {journal} {Classical and Quantum Gravity}\
  }\textbf {\bibinfo {volume} {31}},\ \bibinfo {pages} {082001} (\bibinfo
  {year} {2014})}\BibitemShut {NoStop}%
\bibitem [{\citenamefont {Foo}\ \emph {et~al.}(2021{\natexlab{d}})\citenamefont
  {Foo}, \citenamefont {Arabaci}, \citenamefont {Zych},\ and\ \citenamefont
  {Mann}}]{Foo:2021fno}%
  \BibitemOpen
  \bibfield  {author} {\bibinfo {author} {\bibfnamefont {J.}~\bibnamefont
  {Foo}}, \bibinfo {author} {\bibfnamefont {C.~S.}\ \bibnamefont {Arabaci}},
  \bibinfo {author} {\bibfnamefont {M.}~\bibnamefont {Zych}}, \ and\ \bibinfo
  {author} {\bibfnamefont {R.~B.}\ \bibnamefont {Mann}},\ }\href@noop {} {\
  (\bibinfo {year} {2021}{\natexlab{d}})},\ \Eprint
  {http://arxiv.org/abs/2111.13315} {arXiv:2111.13315 [gr-qc]} \BibitemShut
  {NoStop}%
\bibitem [{\citenamefont {Bekenstein}(1973)}]{bekensteinPhysRevD.7.2333}%
  \BibitemOpen
  \bibfield  {author} {\bibinfo {author} {\bibfnamefont {J.~D.}\ \bibnamefont
  {Bekenstein}},\ }\href {\doibase 10.1103/PhysRevD.7.2333} {\bibfield
  {journal} {\bibinfo  {journal} {Phys. Rev. D}\ }\textbf {\bibinfo {volume}
  {7}},\ \bibinfo {pages} {2333} (\bibinfo {year} {1973})}\BibitemShut
  {NoStop}%
\bibitem [{\citenamefont {Bekenstein}(2020)}]{bekenstein2020quantum}%
  \BibitemOpen
  \bibfield  {author} {\bibinfo {author} {\bibfnamefont {J.~D.}\ \bibnamefont
  {Bekenstein}},\ }in\ \href@noop {} {\emph {\bibinfo {booktitle} {JACOB
  BEKENSTEIN: The Conservative Revolutionary}}}\ (\bibinfo  {publisher} {World
  Scientific},\ \bibinfo {year} {2020})\ pp.\ \bibinfo {pages}
  {331--334}\BibitemShut {NoStop}%
\bibitem [{\citenamefont {Regge}\ and\ \citenamefont
  {Wheeler}(1957)}]{reggePhysRev.108.1063}%
  \BibitemOpen
  \bibfield  {author} {\bibinfo {author} {\bibfnamefont {T.}~\bibnamefont
  {Regge}}\ and\ \bibinfo {author} {\bibfnamefont {J.~A.}\ \bibnamefont
  {Wheeler}},\ }\href {\doibase 10.1103/PhysRev.108.1063} {\bibfield  {journal}
  {\bibinfo  {journal} {Phys. Rev.}\ }\textbf {\bibinfo {volume} {108}},\
  \bibinfo {pages} {1063} (\bibinfo {year} {1957})}\BibitemShut {NoStop}%
\bibitem [{\citenamefont {Hod}(1998)}]{hodPhysRevLett.81.4293}%
  \BibitemOpen
  \bibfield  {author} {\bibinfo {author} {\bibfnamefont {S.}~\bibnamefont
  {Hod}},\ }\href {\doibase 10.1103/PhysRevLett.81.4293} {\bibfield  {journal}
  {\bibinfo  {journal} {Phys. Rev. Lett.}\ }\textbf {\bibinfo {volume} {81}},\
  \bibinfo {pages} {4293} (\bibinfo {year} {1998})}\BibitemShut {NoStop}%
\bibitem [{\citenamefont {Kwon}\ and\ \citenamefont {Nam}(2010)}]{Kwon_2010}%
  \BibitemOpen
  \bibfield  {author} {\bibinfo {author} {\bibfnamefont {Y.}~\bibnamefont
  {Kwon}}\ and\ \bibinfo {author} {\bibfnamefont {S.}~\bibnamefont {Nam}},\
  }\href {\doibase 10.1088/0264-9381/27/12/125007} {\bibfield  {journal}
  {\bibinfo  {journal} {Classical and Quantum Gravity}\ }\textbf {\bibinfo
  {volume} {27}},\ \bibinfo {pages} {125007} (\bibinfo {year}
  {2010})}\BibitemShut {NoStop}%
\bibitem [{\citenamefont {Saravani}\ \emph {et~al.}(2016)\citenamefont
  {Saravani}, \citenamefont {Aslanbeigi},\ and\ \citenamefont
  {Kempf}}]{Saravani_2016}%
  \BibitemOpen
  \bibfield  {author} {\bibinfo {author} {\bibfnamefont {M.}~\bibnamefont
  {Saravani}}, \bibinfo {author} {\bibfnamefont {S.}~\bibnamefont
  {Aslanbeigi}}, \ and\ \bibinfo {author} {\bibfnamefont {A.}~\bibnamefont
  {Kempf}},\ }\href {\doibase 10.1103/physrevd.93.045026} {\bibfield  {journal}
  {\bibinfo  {journal} {Physical Review D}\ }\textbf {\bibinfo {volume} {93}}
  (\bibinfo {year} {2016}),\ 10.1103/physrevd.93.045026}\BibitemShut {NoStop}%
\bibitem [{\citenamefont {Zych}\ \emph {et~al.}(2019)\citenamefont {Zych},
  \citenamefont {Costa}, \citenamefont {Pikovski},\ and\ \citenamefont
  {Brukner}}]{zych2019}%
  \BibitemOpen
  \bibfield  {author} {\bibinfo {author} {\bibfnamefont {M.}~\bibnamefont
  {Zych}}, \bibinfo {author} {\bibfnamefont {F.}~\bibnamefont {Costa}},
  \bibinfo {author} {\bibfnamefont {I.}~\bibnamefont {Pikovski}}, \ and\
  \bibinfo {author} {\bibfnamefont {C.}~\bibnamefont {Brukner}},\ }\href
  {\doibase 10.1038/s41467-019-11579-x} {\bibfield  {journal} {\bibinfo
  {journal} {Nature Communications}\ }\textbf {\bibinfo {volume} {10}}
  (\bibinfo {year} {2019}),\ 10.1038/s41467-019-11579-x}\BibitemShut {NoStop}%
\bibitem [{\citenamefont {Howl}\ \emph {et~al.}(2022)\citenamefont {Howl},
  \citenamefont {Akil}, \citenamefont {Kristj\'ansson}, \citenamefont {Zhao},\
  and\ \citenamefont {Chiribella}}]{Howl:2022oqz}%
  \BibitemOpen
  \bibfield  {author} {\bibinfo {author} {\bibfnamefont {R.}~\bibnamefont
  {Howl}}, \bibinfo {author} {\bibfnamefont {A.}~\bibnamefont {Akil}}, \bibinfo
  {author} {\bibfnamefont {H.}~\bibnamefont {Kristj\'ansson}}, \bibinfo
  {author} {\bibfnamefont {X.}~\bibnamefont {Zhao}}, \ and\ \bibinfo {author}
  {\bibfnamefont {G.}~\bibnamefont {Chiribella}},\ }\href@noop {} {\  (\bibinfo
  {year} {2022})},\ \Eprint {http://arxiv.org/abs/2203.05861} {arXiv:2203.05861
  [quant-ph]} \BibitemShut {NoStop}%
\end{thebibliography}%


\end{document}